\documentclass[epj]{webofc}
\usepackage[varg]{txfonts}   

\usepackage{color}
\definecolor{viv}{RGB}{255,109,254}

\woctitle{QCD@Work 2016}
\begin{document}
\selectlanguage{english}
\title{Gluon fusion contribution to $HBB$ ($B = H,\gamma, Z$) at the LHC}

\author{\firstname{Ambresh} \lastname{Shivaji}\inst{1}\fnsep\thanks{\email{ambresh.shivaji@pv.infn.it}} \and
        \firstname{Pankaj} \lastname{Agrawal}\inst{2} \and
        \firstname{Debashis} \lastname{Saha}\inst{2}
}


\institute{INFN, sezione di Pavia, Via Agostino Bassi, 6-27100 Pavia (PV) Italy 
\and
           Institute of Physics, P.O.: Sainik School, Bhubaneswar - 751005, INDIA
          }

\abstract{
We have calculated one-loop amplitudes for the production of Higgs boson in association 
with two electroweak bosons ($H,\gamma,Z$) via gluon-gluon fusion. 
We present preliminary results for the total cross section at 8, 13 and 100 TeV center-of-mass 
energies at $pp$ colliders. We study the interference effect and, also comment 
on the effect of new physics in terms of anomalous couplings of the Higgs boson in these processes.
}
\maketitle
\section{Introduction}
\label{intro}

There are many standard model (SM) decay or scattering processes which begin at 
loop-level at the leading order itself. Such loop-induced SM processes 
are expected to be sensitive to new physics scales. For example, new heavy particles 
which cannot be produced directly can contribute in the loop in these processes leading 
to a deviation from the SM predictions. We are particularly interested in loop-induced 
gluon fusion processes, which can be  
important at high energy hadron colliders such as the LHC and its future upgrades due 
to the availability of a large gluon flux.

In the past we have studied ${gg \to VVj,~ VHj~(V=\gamma,Z,W)}$ processes at the 
LHC~\cite{Agrawal:2012df,Agrawal:2012as,Agrawal:2014tqa}. In this talk we will focus 
on ${gg \to H B B~(B=H,\gamma,Z)}$ processes. Some results on these processes are reported 
in~\cite{Plehn:2005nk,Binoth:2006ym,Mao:2009jp,Maltoni:2014eza,Hirschi:2015iia,
Papaefstathiou:2015paa}.
Observing ${HHH}$ process would provide us direct information on quartic self-Higgs coupling, while
${HHZ}$ is a background to ${HHH}$ in $HHb{\bar b}$ channel. On the other hand, 
${HVV~(V=\gamma,Z,W)}$ processes are backgrounds to ${gg \to HH}$ when one of the two 
Higgs bosons decays into a pair of vector bosons (${\gamma\gamma,~\gamma Z,~ ZZ^*,~ WW^*}$). 

\begin{figure}[h]
\centering
\includegraphics[width=6cm,clip]{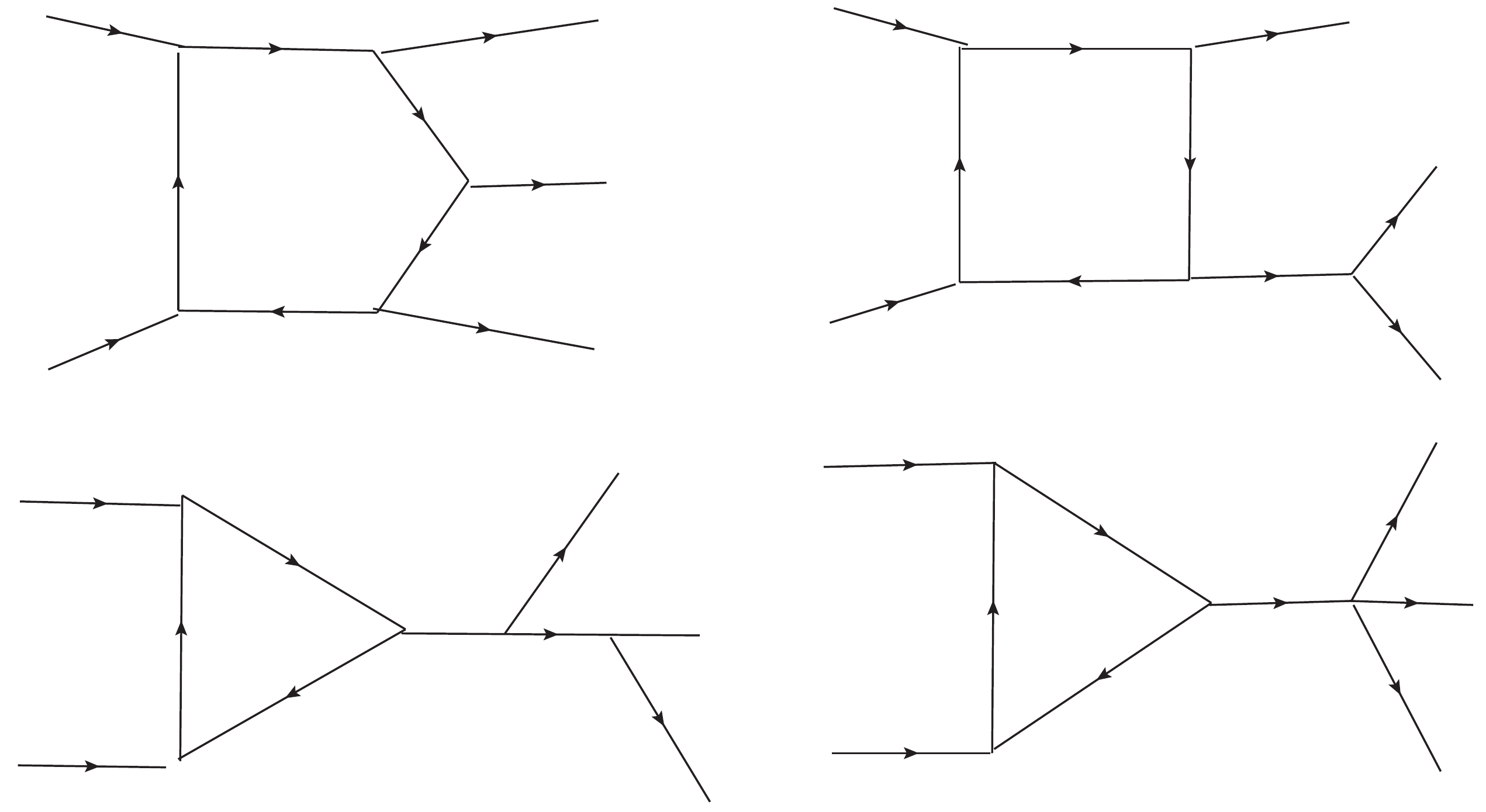}
\caption{Pentagon, box and triangle topologies contributing to ${ gg \to HBB}$ amplitude. In case 
of $H\gamma\gamma$ only pentagon contributes, while $HZ\gamma$ does not receive any triangle contribution.}
\label{fig-1}       
\end{figure}

The processes under consideration are one-loop at the leading order and proceed via quark 
loop diagrams. We have triangle, box and pentagon one-loop amplitudes which contribute to them. 
The one-loop topologies involved are shown in figure~\ref{fig-1}. In most cases, we can identify 
prototype amplitudes and generate all other amplitudes by permuting the external momenta and 
polarizations. Various symmetries can also be utilized to simplify such complex calculations. For example, 
due to charge conjugation, ${\cal M}({gg \to HH\gamma}) = 0$. For the same reason in 
${gg \to HHZ}$ case only the axial-vector part of ${qqZ}$ coupling contributes, while 
in ${gg \to HZ\gamma}$ and ${gg \to HZZ}$ cases only the vector type of amplitude 
gives non-zero contribution.

\section{Calculation and checks}
\label{sec-2}

 We have calculated the quark loop traces in {\tt FORM}~\cite{Vermaseren:2000nd} in 
 $4-$dimensions. Except top quark, all other quarks are considered massless.  
 The $HBB$ amplitude at this stage is expressed in terms of various tensor integrals. 
 One of the most difficult parts of one-loop amplitude calculations is the reduction of
 tensor integrals into a suitable set of scalar integrals.
 In our processes, we have one-loop five point tensor integral of rank four as the most
 complicated tensor integral, 
\begin{equation}
 {\cal E}_{\mu\nu\rho\sigma} = \int {d^nl\over {(2\pi)^n}} {l_\mu l_\nu l_\rho l_\sigma 
                               \over {D_0 D_1 D_2 D_3 D_4 }}.
\end{equation}
Using 4-dimensional {\it Schouten Identity}, we reduce pentagon tensor and scalar integrals 
into lower rank box tensor and scalar integrals. For example, the pentagon scalar 
integral can be written as a linear combination of five box scalar integrals~\cite{vanNeerven:1983vr,Shivaji:2013cca}. 
 Reduction of box and lower tensor integrals into appropriate set of scalar integrals 
is done numerically using the methods of Oldenborgh and Vermaseren ({\tt OV})~\cite{vanOldenborgh:1989wn,Shivaji:2013cca} 
in $n~(=4-2\epsilon)$-dimensions. Finally, all the required scalar integrals are calculated using the 
{\tt OneLOop} package~\cite{vanHameren:2010cp}. 

As we expect, the one-loop $HBB$ amplitude is both ultraviolet ($UV$) and infrared ($IR$)
finite. This is an important check on our calculation. As an ultimate check,
we also check the gauge invariance of the amplitude with respect to the gauge currents. 
This check is done numerically by replacing the polarizations with their respective 4-momenta, 
${\epsilon_\mu(k)\to k_\mu}$ for a given phase space point. We calculate the amplitude
numerically before squaring it to get the cross section.

\section{Numerical results}
\label{sec-3}
We now discuss some preliminary results for our processes at $pp$ colliders.  
In our calculations we use following SM input parameters, 
\begin{eqnarray}
 M_H &=& 125~{\rm GeV},~ M_Z = 91.188~{\rm GeV}, ~ M_W = 80.419~{\rm GeV},\\ \nonumber
 m_t &=& 173~{\rm GeV},
 ~ G_\mu = 1.166389 \times 10^{-5}~{\rm GeV}^{-2}. 
\end{eqnarray}
We also use following basic kinematic cuts to produce results,  
  \begin{eqnarray}
   {\rm p_T^{H/Z}} > 1~{\rm GeV},~ {\rm p_T^\gamma} > 20~{\rm GeV},~  
   |{y_{\rm H/Z}}|< 5.0,~ |{y_\gamma} |< 2.5,~ \Delta {\rm R_{\gamma\gamma}} > 0.4. 
  \end{eqnarray}
Note that the 1 GeV cut on $p_T$ of $H$ and $Z$ is applied mainly to improve the 
numerical stability of the code. Further, we have used {\tt cteq6l1} pdf set~\cite{Nadolsky:2008zw}, and have set the partonic 
center-of-mass energy (cme) as the common scale for renormalization and factorization, 
$\mu_{\rm F} = \mu_{\rm R} = \sqrt{\rm \hat s}$.

\subsection{SM prediction}
\label{sec-3.1}
\begin{figure}[h]
\centering
  \includegraphics[width=6cm,clip]{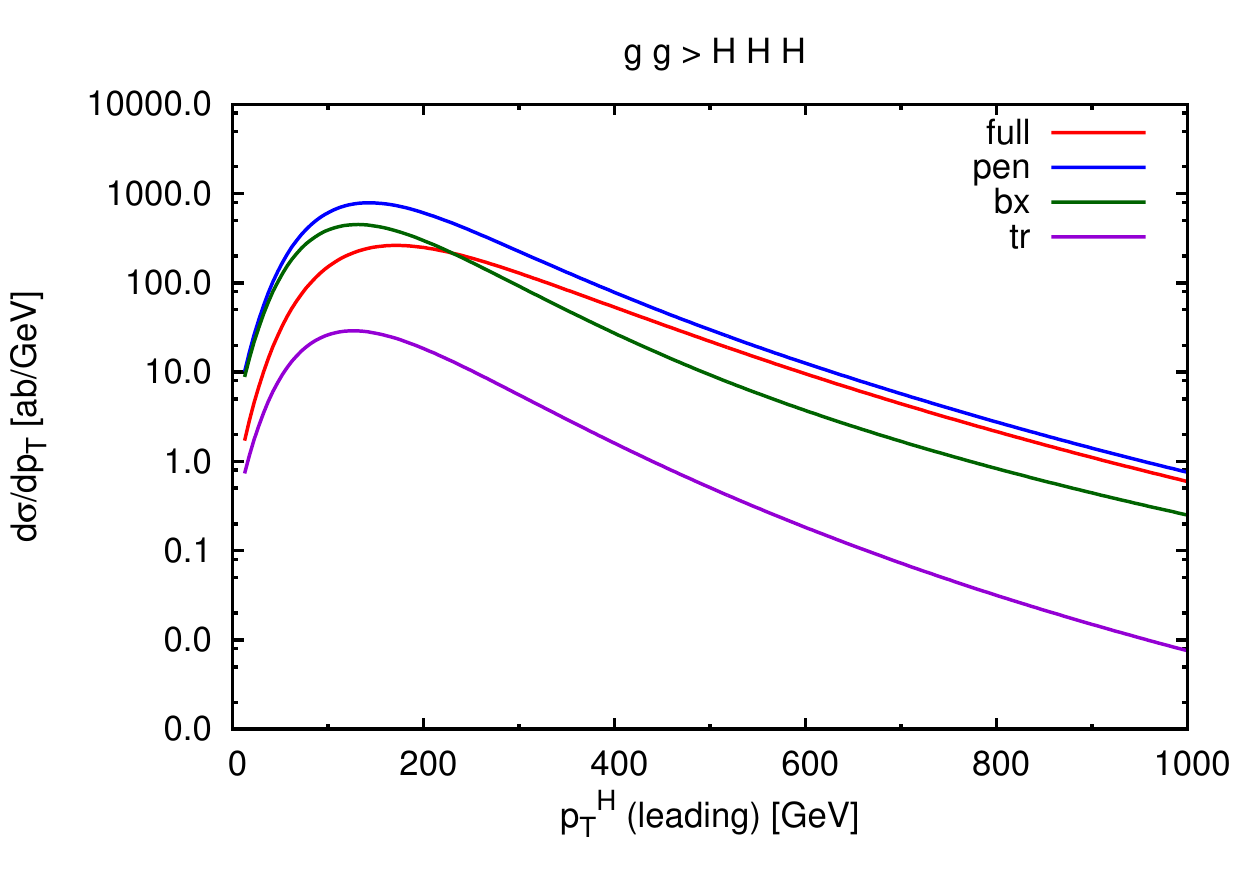}
  \includegraphics[width=6cm,clip]{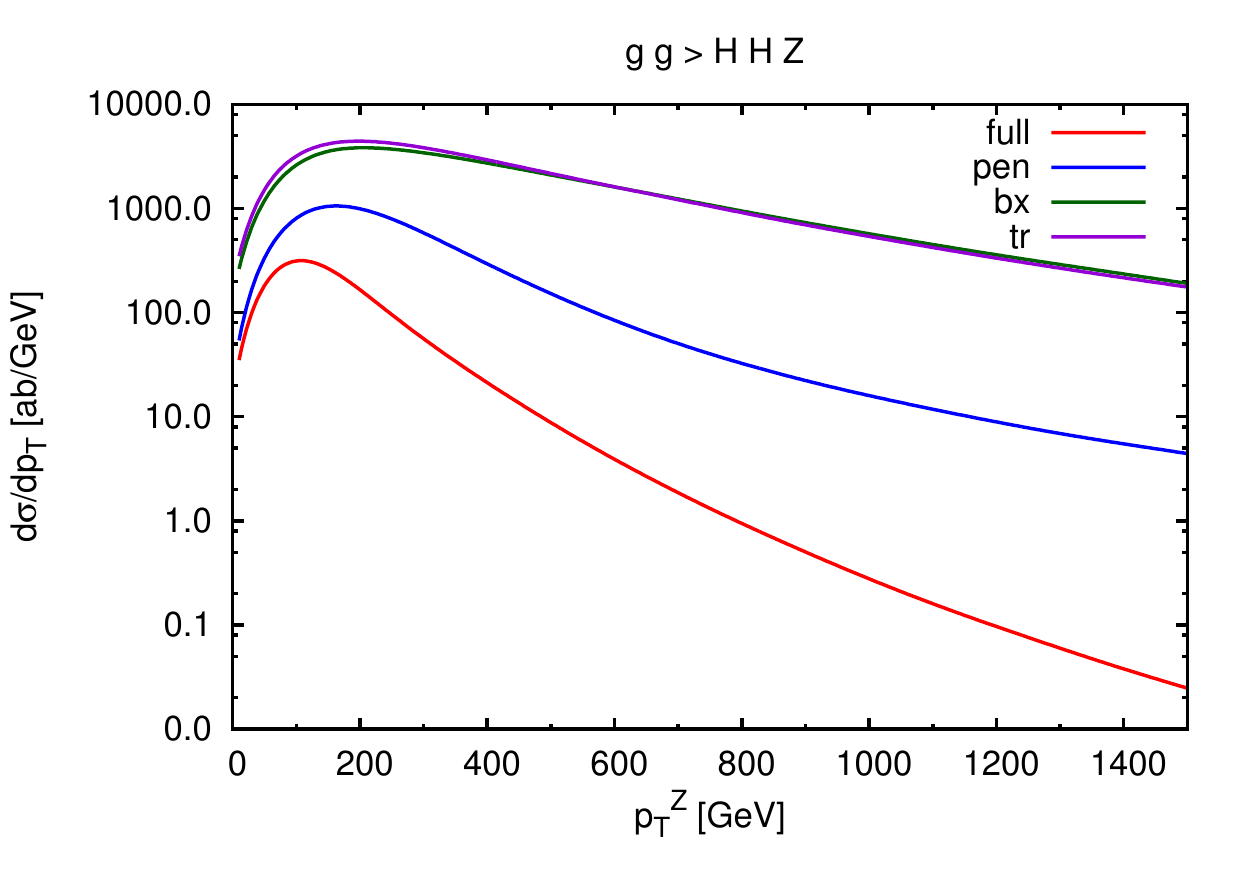}
  \includegraphics[width=6cm,clip]{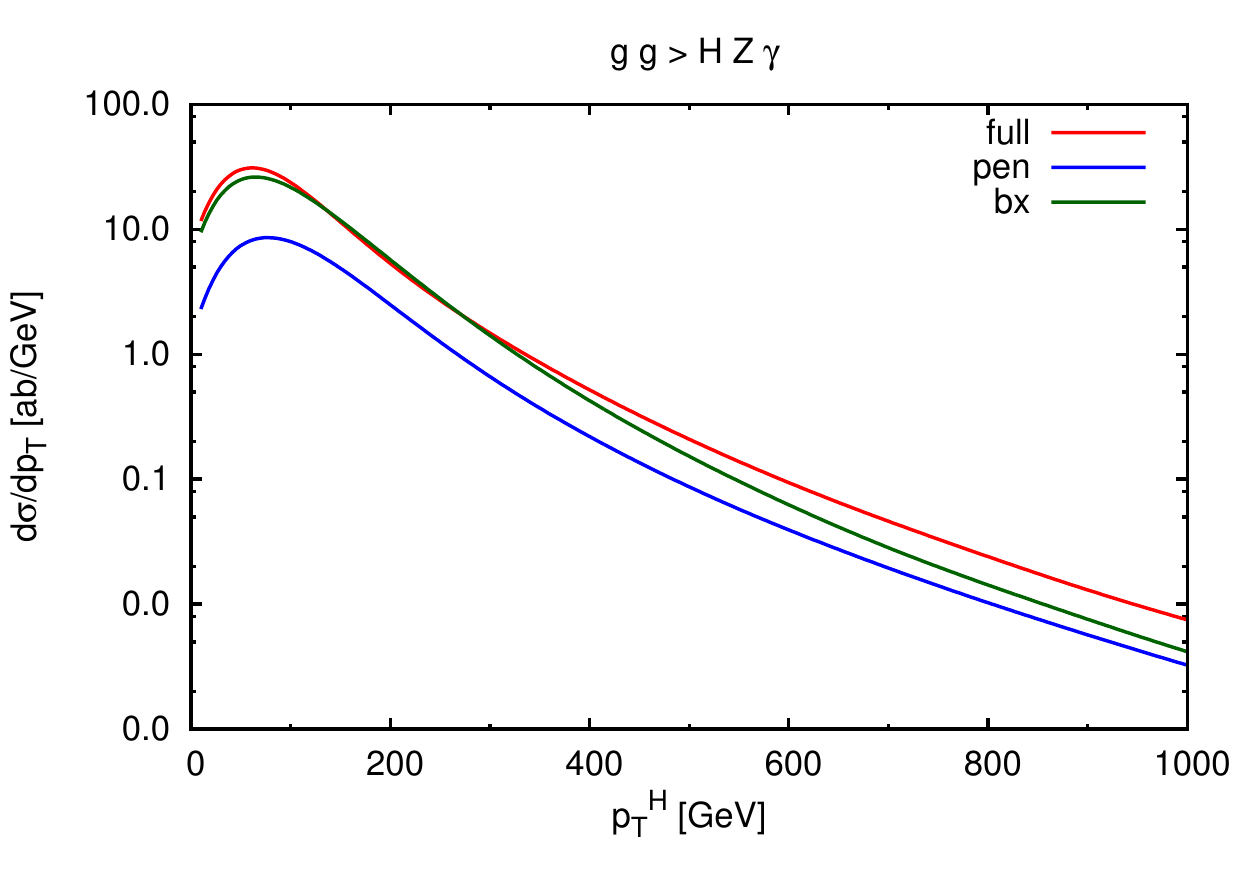}
  \includegraphics[width=6cm,clip]{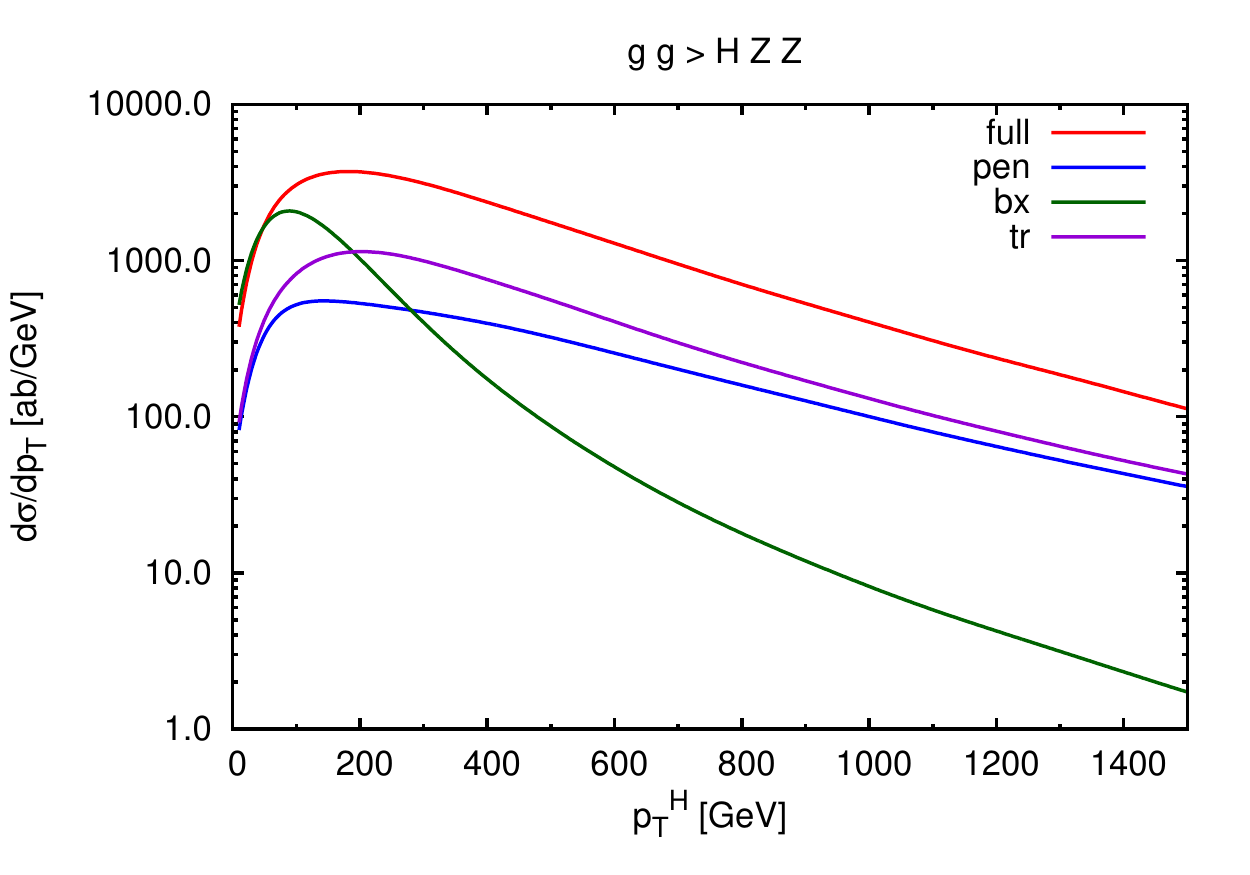}
\caption{A comparison among various pieces of the amplitude at the level of 
kinematic distributions in $gg \to HBB$ processes.}
\label{fig-2}       
\end{figure}

In table~\ref{tab-1}, we report hadronic cross sections for ${gg \to HBB}$ processes 
at various collider center-of-mass energies. We also mention the percentage scale uncertainties 
when the scale is changed by a factor of two around its central value. Due to heavy particles 
in the final state and presence of many electroweak couplings, these processes have very small 
cross sections even at 100 TeV.
It should be noted that the triangle, box and pentagon amplitudes are separately gauge
invariant {\it with respect to the gluons} in all the processes. To understand the interference effect
among these amplitudes, we have computed their individual contributions at the cross section 
level in table~\ref{tab-2}. 
It should be kept in mind that only the full contribution is meaningful and consistent with the complete SM symmetry.
We note that except in $HZZ$ case, in all other cases there is a destructive interference between 
amplitudes. This destructive interference is weakest in $HZ\gamma$ while strongest in $HHZ$. 
The $HZZ$ amplitude displays a very strong constructive interference. Similar feature 
we observe in $HWW$, however, our calculation is not yet complete~\cite{Agrawal:2016xxx}.
In figure~\ref{fig-2} we have selected some kinematic distributions to 
highlight the variation of the interference effect between amplitudes with respect to 
a scale like $p_T$.

\begin{table}
\centering
\caption{SM cross sections at various collider center-of-mass energies with scale uncertainties. 
All cross sections are in $ab$.}
\label{tab-1}       
\begin{tabular}{llll}
  \hline
   $\sqrt{\rm s}$~(TeV)       & 8       & 13       & 100 \\
  \hline
        \\
   $\sigma$ ($HHH$)             & $7.048^{+34\%}_{-24\%}$   & $31.87^{+30\%}_{-22\%}$    & $3093^{+17\%}_{-14\%}$ \\
  \hline 
        \\
   $\sigma$ ($ HHZ$)             &  $10.11^{+34\%}_{-24\%}$  &  $42.76^{+30\%}_{-22\%}$  & $3468^{+17\%}_{-14\%}$  \\
  \hline
        \\
   $\sigma$ ($ H\gamma\gamma $)  &  $1.240^{+37\%}_{-23\%}$  &  $4.852^{+29\%}_{-22\%}$   & $265.8^{+16\%}_{-13\%}$ \\
  \hline
        \\
   $\sigma$ ($ HZ\gamma $)       &  $1.401^{+32\%}_{-22\%}$  &   $4.931^{+28\%}_{-21\%}$  &  $241.3^{+15\%}_{-13\%}$\\
  \hline
        \\
   $\sigma$ ($ HZZ$)             & $83.7000^{+36\%}_{-21\%}$   &  $471.636^{+36\%}_{-24\%}$   & $102573^{+20\%}_{-15\%}$ \\
\end{tabular}
\end{table}

\begin{table}
\centering
\caption{Contributions from pentagon, box and triangle amplitudes at $\sqrt{\rm s}$ = 100 TeV. 
          All cross sections are in $ab$.}
\label{tab-2}       
\begin{tabular}{lllll}
  \hline
     & PEN      & BX       & TR    & FULL   \\
  \hline
   $\sigma$ ($HHH$)            & 8110    & 4319    & 274.2   & 3039 \\
  \hline 
   $\sigma$ ($ HHZ$)            & 17214.5 & 116996  & 125552  & 3468.37 \\
  \hline
  $\sigma$ ($ H\gamma\gamma$)   & 265.8   & --      & --      & 265.8 \\
  \hline
   $\sigma$ ($ HZ\gamma$)       & 78.04   & 216.2   & --      & 241.3 \\
  \hline
   $\sigma$ ($ HZZ$)            & 18677.3 & 23684.9 & 31998.7 & 102573 \\
\end{tabular}
\end{table}

\subsection{Higgs Anomalous couplings}
\label{sec-3.2}
New physics beyond the SM can induce modifications to couplings among SM particles. The couplings 
of the Higgs boson with heavy fermions, massive gauge bosons and the Higgs self couplings are of 
particular interest in this regard. To demonstrate the effect of modifying Higgs couplings in 
our processes, we have scaled the SM Higgs couplings by factor $C_i~(i=ttH,~ZZH,~ZZHH,~3H,~4H)$. 
In table~\ref{tab-3} we report the percentage deviation in total cross section when $C_i$ 
is changed by $\pm 10\%$ from its SM value ($C_i^{\rm SM}=1$), which is consistent with current 
LHC data on Higgs. We modify one coupling at a time which indeed provides valuable information 
on its role in changing the interference pattern among various diagrams. 
The table can be more or less understood with the help of table~\ref{tab-2} which has 
the information on triangle, box and pentagon contributions which in presence of anomalous 
couplings depend on $C_i$ in a specific manner. 
We can see that some of our processes are quite sensitive to modifications in $ttH$ and 
$ZZH$ couplings. Note that 
we have ignored correlations among the couplings which may arise in a given model of new physics 
or in presence of higher dimensional operators introduced to capture new physics~\cite{Agrawal:2016xxx}.

\begin{table}
\centering
\caption{The effect of changing various couplings by $10\%$ of their SM values at $\sqrt{\rm s}$ = 100 TeV. 
           The first and second entries in each parentheses correspond 
          to $C_i = 0.9$ and $1.1$ respectively.}
\label{tab-3}       
\begin{tabular}{llllll}
  \hline
   {\bf ANML}\\(0.9,1.1)  & {\large $C_{\rm ttH}$} & {\large $C_{\rm 3H}$} & {\large $C_{\rm 4H}$} 
                            & {\large $C_{\rm ZZH}$} & {\large $C_{\rm ZZHH}$} \\
  \hline \\
   $HHH$             & ($-52\%,+92\%$)   & ($+8\%,-5\%$)     & ($+1\%,-1\%$)  & --  & -- \\
  \hline \\
   $ HHZ$             & ($+22\%,+81\%$)   & ($-1\%,+1\%$) & -- & ($+127\%,+140\%$) & ($-6.3\%,+18\%$) \\
  \hline \\
   $ H\gamma\gamma$   & ($-1\%,+1\%$)     &  --   & -- & -- & -- \\
  \hline \\
   $ HZ\gamma $       & ($-4\%,+4\%$)     &  --   & -- & ($-15\%,+15\%$) & --\\
  \hline \\
   $ HZZ$             & ($-21\%,+25\%$)   & ($+0.5\%,-0.4\%$)  & -- & ($-26\%,+34\%$) & ($+4\%,-3\%$) \\
\end{tabular}
\end{table}

\section{Conclusion}
\label{sec-4}

We have computed loop-induced gluon fusion contributions to 
${HBB~ (B=H,\gamma,Z)} $ processes at $pp$ colliders. 
One would like to observe $HHH$ to probe 
the quartic self-coupling of the Higgs boson. Others are also backgrounds to 
double Higgs production which carries the direct information on the 
trilinear self-coupling of the Higgs boson. 
We find that due to small rates, their observation would require a very 
large luminosity. 
Some of these processes display a strong interference between different 
classes of diagrams. We have seen that any modification to the SM Higgs couplings due to
new physics effects can spoil the interference and lead to a very different prediction.
The effect of anomalous couplings can be studied more systematically using higher 
dimension operators which would inherently take care of possible correlations among 
various Higgs couplings.


%
%
%

\end{document}